# AI Ethics:
## A Bibliometric Analysis, Critical Issues, and Key Gaps

Di Kevin Gao, California State University, East Bay, USA*
https://orcid.org/0009-0008-7391-5208

Andrew Haverly, Mississippi State University, USA

Sudip Mittal, Mississippi State University, USA
https://orcid.org/0000-0001-9151-8347

Jiming Wu, California State University, East Bay, USA

Jingdao Chen, Mississippi State University, USA
https://orcid.org/0000-0002-5133-9552

## ABSTRACT

Artificial intelligence (AI) ethics has emerged as a burgeoning yet pivotal area of scholarly research. This study conducts a comprehensive bibliometric analysis of the AI ethics literature over the past two decades. The analysis reveals a discernible tripartite progression, characterized by an incubation phase, followed by a subsequent phase focused on imbuing AI with human-like attributes, culminating in a third phase emphasizing the development of human-centric AI systems. After that, they present seven key AI ethics issues, encompassing the Collingridge dilemma, the AI status debate, challenges associated with AI transparency and explainability, privacy protection complications, considerations of justice and fairness, concerns about algocracy and human enfeeblement, and the issue of superintelligence. Finally, they identify two notable research gaps in AI ethics regarding the large ethics model (LEM) and AI identification and extend an invitation for further scholarly research.

## INTRODUCTION

In November 2022, OpenAI took the world by storm with the debut of ChatGPT. The subsequent release of Bard by Google in February 2023 opened the floodgate of the once carefully guarded AI underworld. It also laid bare the upcoming breakneck competition that will reshuffle the winners and losers at the pivotal moment of Industry 4.0 (McKinsey, 2022), where a trivia chat-bot mistake cost

*Corresponding Author

Google $100 billion in market capitalization (Wittenstein, 2023) and a tiny graphics chip propelled NVIDIA into the trillion-dollar club (Fitch, 2023). Tech giants such as Apple, Amazon, and Facebook swiftly joined the race (Gurman, 2023; Dotan, 2023; Hao, 2023). The burgeoning excitement for anything AI also fueled a powerful surge in AI startups, with Anthropic AI and Inflection AI becoming newly minted unicorns despite being in the business for less than two years (Hu & Shekhawat, 2023; Konrad, 2023). New AI startups flourished.

The study of AI ethics is the study of the ethical and responsible development and deployment of artificial intelligence technology. Its significance is underscored by the rapid advancements in AI technology and the potential disruptions it may bring to our society. However, crucial questions need to be answered: How has AI ethics evolved, and what are the critical issues and debates? Additionally, what are the key gaps that necessitate further scholarly research? Our research is built on prolific AI ethics literature published between 2004 and 2023, a span of 20 years. Utilizing keyword patterns, we systematically analyze the development phases and trends in AI ethics. Drawing from comprehensive literature reviews, we present seven key issues that continue to be subjects of research and debate. Finally, we extend an invitation for additional scholarly research on the large ethics model (LEM) and AI identification.

This article provides a distinctive contribution to AI ethics across four areas:

- The delineation of the origins of modern AI ethics
- The contrast of human-like machines versus human-centric machines, highlighting two pivotal phases in AI development
- The initiation of the LEM discussion

AI ethics research may leverage the approaches used by the large language model (LLM) and get away from the bounds of conventional approaches of theories, principles, and frameworks.

- The initiation of discussions on AI identification

AI ethicists can remove mysteries and nebulosity on AI by pioneering approaches to identify and rate AI instances.

## DEFINITIONS

Artificial intelligence (AI) was coined in 1955 by John McCarthy, Marvin Minsky, Nathaniel Rochester, and Claude Shannon during the preparation for the Dartmouth Workshop (Dartmouth, 1956). John McCarthy defined AI as:

*The science and engineering of making intelligent machines, especially intelligent computer programs. It is related to the similar task of using computers to understand human intelligence, but AI does not have to confine itself to methods that are biologically observable. (2007, p. 2)*

The term *ethics* originates from the Greek word "ethos" meaning "character." In the field of philosophy, ethics is the field that investigates individual behavior in society, providing rational justification for moral judgments, discerning what is morally right or wrong, and distinguishing what is morally just or unjust (Cornell, 2023). Artificial intelligence ethics is the study of rational justifications, what is morally right or wrong and just or unjust, for the responsible development and deployment of artificial intelligence technology.

# METHOD

This literature review combines bibliometric analysis and a selected literature walk-through, utilizing SCOPUS as the primary data source supplemented by Google Scholar and Mendeley. We used VOSviewer as the principal data aggregator for keyword analysis. The detailed methodology is described in the following sections.

## Literature Collection for Bibliometric Analysis of the Origin of AI Ethics

We searched the SCOPUS database for "AI ethics" OR "artificial intelligence ethics" OR "machine ethics" OR "algorithm ethics" OR "information ethics" OR "ethics of technology" OR "Robotic Ethics" OR "Robot Ethics" OR "artificial moral agent" OR "artificial moral agents" between 2004 and 2023, a period of 20 years across all languages, countries, and territories. A total of 2,517 articles were initially identified. We scanned the full list and removed 59 entries that missed important authorship information and one entry that missed sources. The refined dataset, comprising 2,457 articles, was used to calculate AI ethics keyword frequencies in Section 4 - the Bibliometric Analysis of the Inception of AI ethics.

## Literature Collection for Keyword Analysis of AI Development Phases

In the subsequent phase of literature collection, the SCOPUS database was revisited to specifically search for articles using the following keywords: "AI ethics" OR "Artificial intelligence ethics" OR "co-existence of 'Ethics' and 'Artificial Intelligence'" OR "co-existence of 'Ethical' and 'Artificial Intelligence.' This yielded a selection of 759 articles that formed the basis for the keyword analysis presented in the results section, which delves into the developmental phases of AI ethics based on keyword patterns.

During the analysis, co-occurrence author keywords from VOSviewer were employed, utilizing the "full counting" method where each keyword is weighted the same irrespective of how many keywords are listed in the literature.

## Manual Literature Walk-Through

For the review of key issues, we chose literature with the highest citations and relevance. Throughout this process, we were very selective in the inclusion of literature from specific domains, such as medicine, healthcare, and education. This decision was guided by our intention to reserve discussion of domain-specific ethics for a different article or forum. This selection approach enhances the robustness of the "Key Issues in AI Ethics" section.

# RESULTS

## Bibliometric Analysis of the Inception of AI Ethics

Early ethical studies in artificial intelligence trace back to Isaac Asimov in 1940, who presented his three laws of robotics in the short story "Runaround" (Moran, 2008), predating the formalization of AI ethics as an academic discipline. In the subsequent decades, there were sporadic discussions of machine intelligence ethics, which precede AI ethics. To understand AI ethics and its origins, an exploration of AI's historical context and its development is essential.

As noted in the "Definitions" section, AI was coined in 1955, in preparation for the Dartmouth Workshop. Since its inception, AI has experienced multiple boom and bust cycles. It garnered significant attention from its outset until 1973. In 1974, however, the British Lighthill Report triggered a substantial loss of confidence in AI within the academic establishment in the United Kingdom (Lighthill, 1973). Coupled with funding cuts at the Defense Advanced Research Projects Agency (DARPA) in the United States, AI entered an "AI winter" until 1980. Following a resurgence that lasted seven years, AI faced another setback in 1987 mainly due to the collapse of Lisp machines and the failure of the expert systems. By the early 2000s, AI struggled with a tarnished reputation due to persistent overpromises and under-deliveries. Some computer scientists and software engineers even began avoiding the term "artificial intelligence" (Markoff, 2005). In the 1990s and 2000s, new computer science disciplines flourished, such as informatics, machine learning, machine perception, analytics, predictive analytics, language models, or computational intelligence. However, they were deliberately not categorized under artificial intelligence.

Figure 1 shows that the term "AI ethics" first appeared in literature keywords in 2008. Before that, AI ethics discussions were dispersed across other related fields such as information ethics, machine ethics, roboethics, technology ethics, and computer ethics. After the first occurrence in 2008, however, AI ethics was in hibernation for the next five years despite significant advancements in AI technology, such as IBM Watson winning the Jeopardy game in 2011 (Hale, 2011). The pivotal year for AI's resurgence was 2012 when AlexNet employed GPU to train its CNN model, winning the ImageNet 2012 Challenge (Krizhevsky et al., 2017; Gershgorn, 2018). This renewed interest in AI led to a surge in funding, prompting researchers to associate their work with AI once again. AI expanded its purview to encompass any computer science discipline that enables human-like intelligence. AI became an aggregator and a destination.

The frequency of AI ethics in literature titles and keywords mirrored the ascendance of AI. Based on Figure 1, AI ethics usage has surged from one occurrence in 2014 to 148 in 2022. In the partial year of 2023 (up to July 28), AI ethics has already appeared 114 times. Notably, AI ethics has established a substantial lead over other related ethics fields. Based on our bibliometric analysis, we believe 2014 marks the formal establishment of AI ethics as an academic discipline. The pre-2014 period is considered AI ethics' incubation period.

## AI ETHICS DEVELOPMENT PHASES BASED ON KEYWORD ANALYSIS

In this section, we analyze the frequency of "AI Ethics" or "Artificial Intelligence Ethics" in literature keywords over the past two decades to unpack trends in the development of AI ethics. As previously noted, 2014 stands out as the pivotal year marking the formal establishment of AI ethics as an academic discipline. Before 2014, AI ethics work was dispersed across diverse domains. We conducted additional research to explore keyword usage patterns, leading to the generation of Figure 2 in VOSviewer. This diagram visually represents keywords associated with AI ethics in the last 20 years, with the size of the bubbles indicating the relative frequency of each keyword.

We chronologically reorganized this information and further bifurcated the keywords based on product orientation and their alignment with AI ethics principles, as depicted in Figure 3. The central portion of Figure 3 denotes when keywords began to be consistently used, with product-oriented keywords positioned at the top and AI ethics principle-related keywords in the middle.

The findings were revealing. Beyond the establishment of AI ethics as an academic discipline in 2014, a discernible shift in AI ethics research principles became evident starting from 2020. In the period spanning 2014 to 2019, AI ethics keywords reflected an excitement to imbue AI with ethical attributes akin to miniature humans, including features like fairness, trust, explainability, responsibility, accountability, and transparency. However, commencing in 2020, the emphasis within AI ethics keywords has progressively pivoted towards ensuring no bias, no discrimination, support for diversity, less opacity, and more trustworthiness. This marked a significant transition towards prioritizing downside protection and driving AI technology with a human-centric focus.

In light of these observations, we propose categorizing the development of AI ethics into the following three distinct phases:

Phase I: Incubation (2004-2013)
Phase II: Making AI Human-Like Machines (2014 to 2019)

**Figure 1. Usage of different ethics in literature keywords**

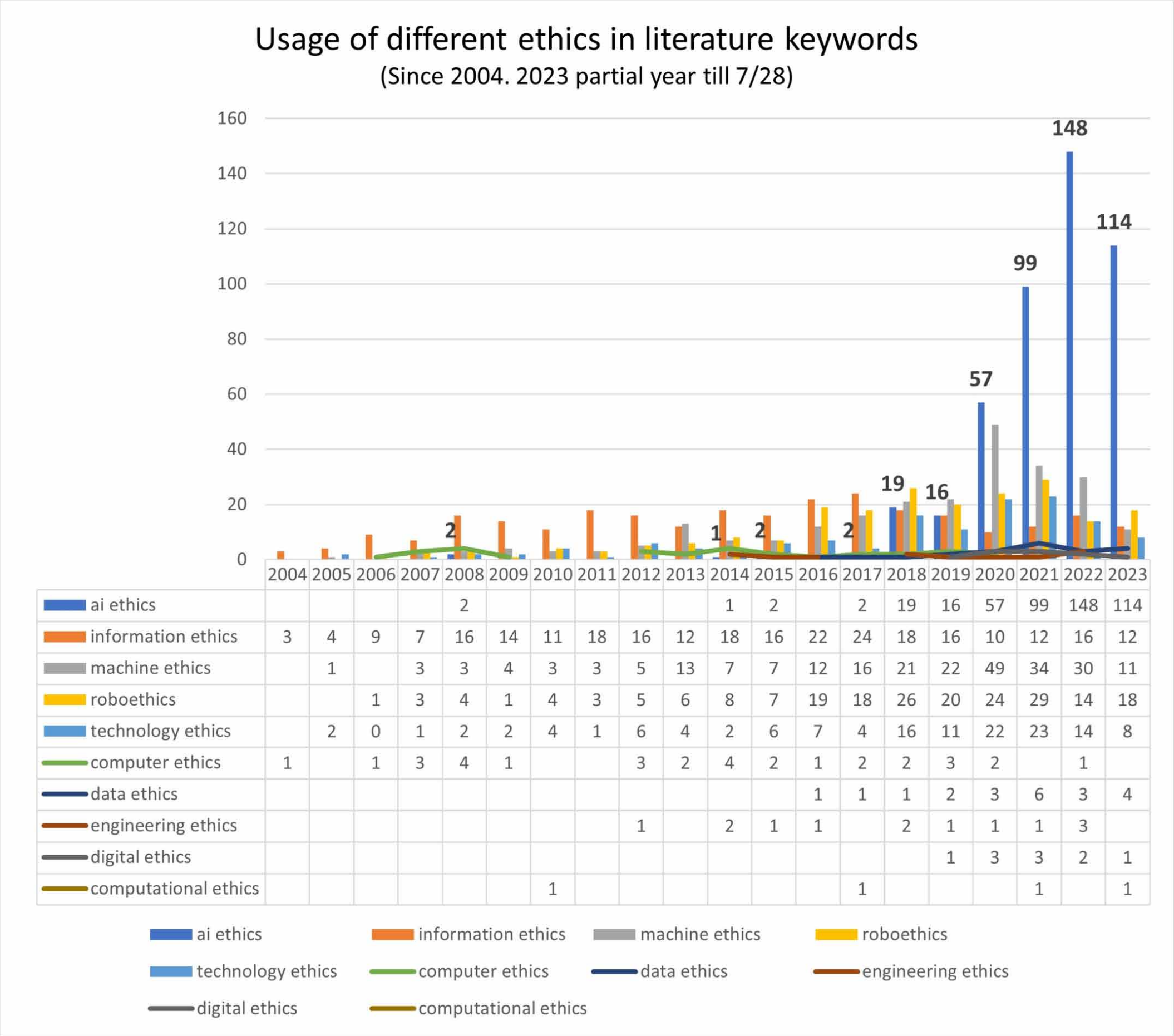

| | 2004 | 2005 | 2006 | 2007 | 2008 | 2009 | 2010 | 2011 | 2012 | 2013 | 2014 | 2015 | 2016 | 2017 | 2018 | 2019 | 2020 | 2021 | 2022 | 2023 |
|---|---|---|---|---|---|---|---|---|---|---|---|---|---|---|---|---|---|---|---|---|
| ai ethics | | | | | 2 | | | | | | 1 | 2 | | 2 | 19 | 16 | 57 | 99 | 148 | 114 |
| information ethics | 3 | 4 | 9 | 7 | 16 | 14 | 11 | 18 | 16 | 12 | 18 | 16 | 22 | 24 | 18 | 16 | 10 | 12 | 16 | 12 |
| machine ethics | | 1 | | 3 | 3 | 4 | 3 | 3 | 5 | 13 | 7 | 7 | 12 | 16 | 21 | 22 | 49 | 34 | 30 | 11 |
| roboethics | | | 1 | 3 | 4 | 1 | 4 | 3 | 5 | 6 | 8 | 7 | 19 | 18 | 26 | 20 | 24 | 29 | 14 | 18 |
| technology ethics | | 2 | 0 | 1 | 2 | 2 | 4 | 1 | 6 | 4 | 2 | 6 | 7 | 4 | 16 | 11 | 22 | 23 | 14 | 8 |
| computer ethics | 1 | | 1 | 3 | 4 | 1 | | | 3 | 2 | 4 | 2 | 1 | 2 | 2 | 3 | 2 | | 1 | |
| data ethics | | | | | | | | | | | | | 1 | 1 | 1 | 2 | 3 | 6 | 3 | 4 |
| engineering ethics | | | | | | | | | 1 | | 2 | 1 | 1 | | 2 | 1 | 1 | 1 | 3 | |
| digital ethics | | | | | | | | | | | | | | | | 1 | 3 | 3 | 2 | 1 |
| computational ethics | | | | | | | 1 | | | | | | | 1 | | | | 1 | | 1 |

Phase III: Making AI Human-Centric Machines (2020 and beyond)

## Phase I: Incubation (2004-2013)

The incubation phase of AI ethics, spanning from 2004 to 2013, was driven by the rapid advancements in artificial intelligence. Pivotal moments such as IBM's Deep Blue defeating world chess champion Garry Kasparov in 1997, Stanford's autonomous vehicle successfully navigating 212 kilometers of off-road terrain in 2005, and IBM's Watson triumphing in Jeopardy! in 2011, captured public attention, portraying AI as both impressive and entertaining.

In 2004, prominent keywords included "personhood," "friendly AI," and "anthropocentrism," illustrating an intense discussion of AI's status and an embryonic state of AI ethics as an academic discipline. AI ethics literature was dispersed across various domains such as information ethics, machine ethics, roboethics, technology ethics, and computer ethics.

During this phase, notable authors, including Michael Anderson, Susan Leigh Anderson, James Moor, Luciano Floridi, and Wendell Wallach, played instrumental roles. Their work, exemplified by Anderson and Anderson's exploration of machine ethics in their research (Anderson et al., 2005), introduced an ethical dimension beyond the traditional focus of computer ethics on human-machine interactions. Machine ethics, a key area of debate, aimed to ensure ethically acceptable behavior of

Figure 2. An analysis of keywords in AI literature from 2004 to 2023 revealed key disciplines in the development of AI ethics

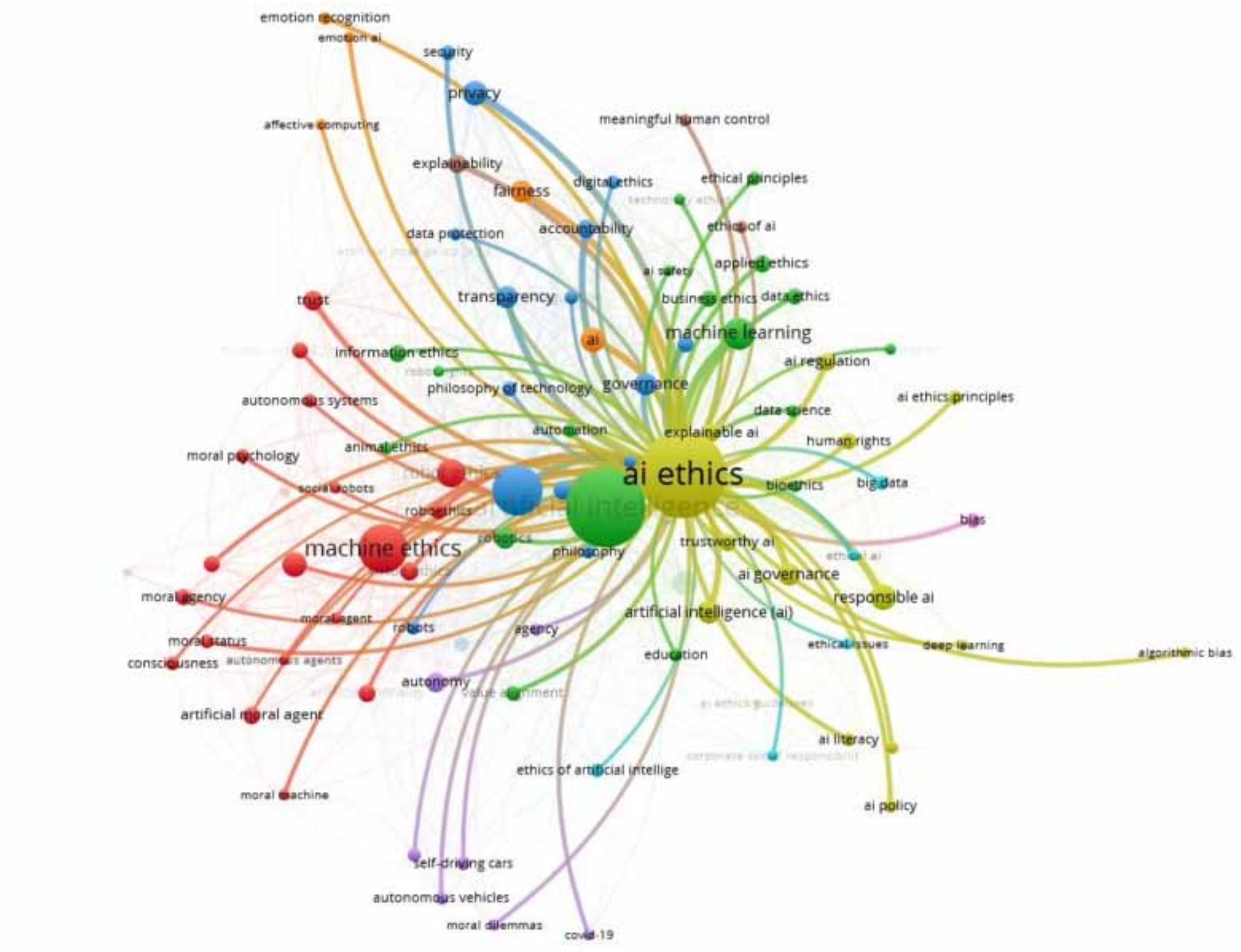

Figure 3. AI ethics development based on keyword analysis

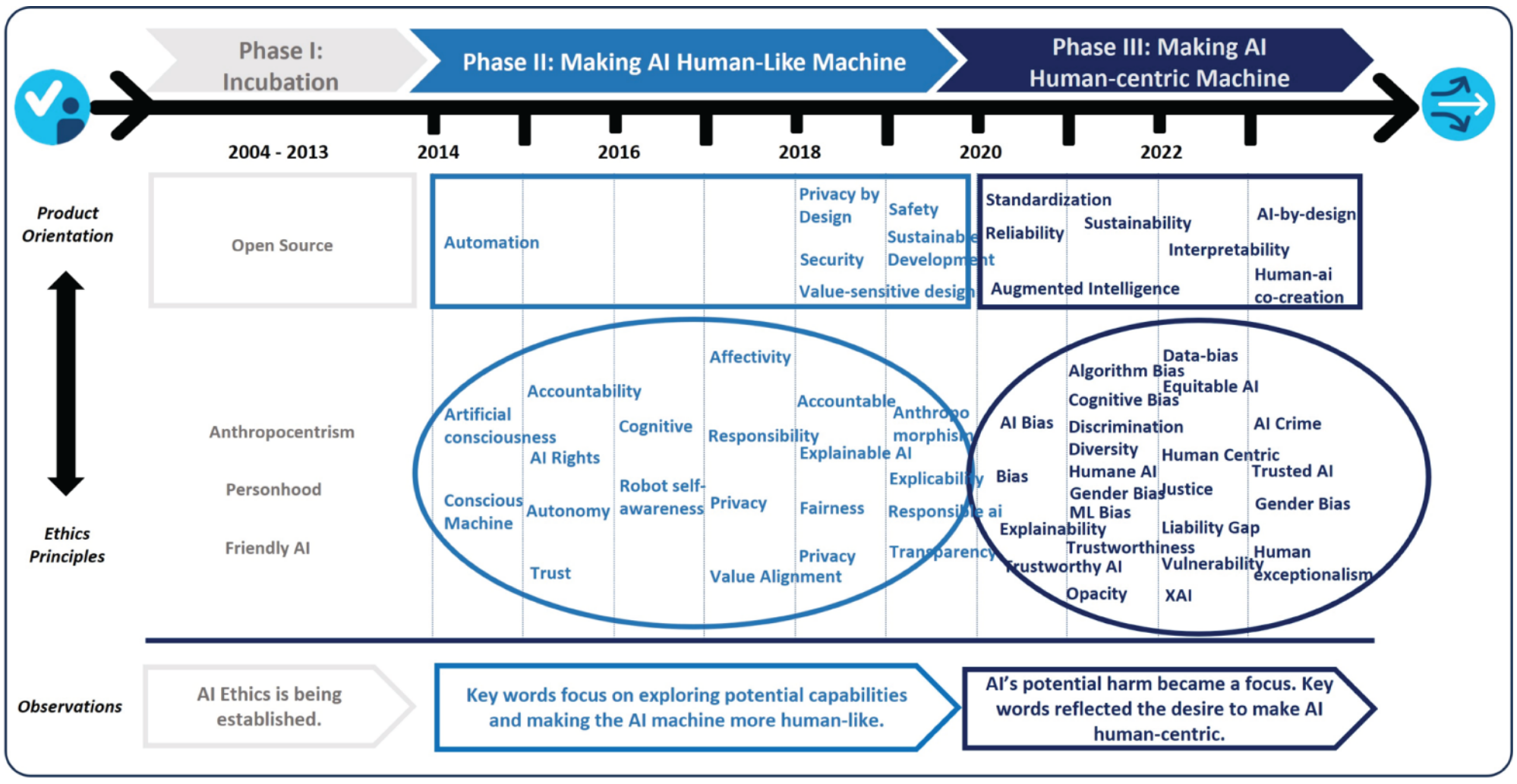

**Note. This diagram is based on AI ethics keywords between 2004 to 2023. Based on our analysis AI ethics experienced three major phases in the last 20 years: Phase I. Incubation (2004-2013); Phase II. Making AI Human-Like Machine (2014-2019); Phase III. Making AI Human-Centric Machine (2020 onwards).**

machines towards human users and other machines. Moor's research challenged the community to develop machines into explicit ethical agents, navigating questions about the status of machines as ethical-impact agents, implicit ethical agents, explicit ethical agents, or full ethical agents (Moor, 2006). Floridi initially focused on informational privacy in the digital age (Floridi, 2005) but delved deeper into the nature and scope of information ethics (Floridi, 2006b). Wallach, Allen, and others posed fundamental questions, including "Why machine ethics?" (Allen et al., 2006) and researched the conceptual and computational models of moral decision-making (Wallach et al., 2010).

These early papers laid a solid foundation for AI ethics as an emerging academic discipline, shifting the research beyond theoretical considerations into practical applications.

### Phase II: Making AI Human-Like Machines (2014-2019)

In Phase II, spanning from 2014 to 2019, AI showcased the potential to function like a human. The focus of AI ethics during this period shifted towards the ethical applications of AI as a mini human across diverse fields. That was the reason we labeled this phase as making AI "human-like" machines.

In 2014, the generative adversarial network (GAN) was developed, enabling the generation of creative images from existing ones. The year 2016 witnessed DeepMind's AlphaGo defeating world Go champion Lee Sedol over five matches (Cho, 2016). By 2018, AI demonstrated superior accuracy in detecting skin cancer compared to dermatologists (Lardieri, 2018). Google Waymo's Robotaxi started services in Phoenix in the same year (Fingas, 2019). These achievements garnered widespread attention, heightened public curiosity and expectations, and positioned AI as a potential mini-human with boundless applications.

Keyword analysis during this period revealed the emergence of new ethical principles featuring human-like attributes, including "accountability," "anthropomorphism," "explicability," "explainable," "fairness," "responsible," "trust," and "transparency." The AI ethics community aimed to imbue this intelligent technology with accountability, explainability, fairness, responsibility, trustworthiness, and transparency, mirroring an ethical human.

Prominent papers during this phase centered on the development of AI ethics frameworks and roadmaps across diverse fields. In 2018, Winfield and Jirotka devised a roadmap linking ethics, standards, regulation, responsible research and innovation, and public engagement to guide ethical governance in robotics and AI (Winfield & Jirotka, 2018). In 2019, Ville et al. presented a research framework for implementing AI ethics in industrial settings (Ville et al., 2019). Boesl and Bode (2019) proposed the adaptation of labeling as a method to foster "Robotic AI Governance," a concept of self-regulation for robotics, automation, and artificial intelligence. LaBrie et al. (2019) provided a framework for ethical audits of AI algorithms.

### Phase III: Making AI Human-Centric (2020-Present)

In Phase III, amid its rapid ascent, AI revealed aspects far from angelic, prompting the AI ethics community to establish disciplines aimed at constraining this powerful force within our delicate societal framework. This phase, titled "Making AI Human-Centric Machines," reflects a concerted effort to ground AI as an unbiased, non-discriminatory, diversity-supporting, less opaque, and more trustworthy servant to humans.

By 2020, AI had surpassed human performance in various domains such as handwriting recognition, speech recognition, image recognition, reading comprehension, and language understanding (Kiela et al., 2021). In 2021, OpenAI's DALL-E could generate high-quality images from written descriptions. Together with Deep Fakes, it exacerbated online misinformation and eroded basic human trust. In July 2022, Google was embroiled in controversy when it fired an engineer who claimed its LaMDA language model was sentient (Kruppa, 2022). In November 2022, OpenAI released Chat GPT 3.0 to the public and triggered an all-out AI race. AI companies competed publicly and fiercely which alarmed the public that ethics and safety would be put on the back burner (Lock, 2022). In 2023, Goldman Sachs estimated that 300 million jobs could be replaced by AI (Nolan, 2023).

Scientists, entrepreneurs, and public officers started to remind the general public of the consequences of unbridled AI development. Public distrust of AI surged.

During this phase, many precautionary keywords showed up, such as “bias” and “AI bias” in 2020, “algorithm bias,” ‘cognitive bias,” “gender bias,” “machine learning bias,” “diversity,” “discrimination,” and “opacity” in 2021; “data bias,” “human-centric,” “equitable AI,” “liability gap,” “justice,” and XAI in 2022; and “gender bias,” “trusted AI,” and “human exceptionalism” in 2023. The ethics community wanted to make AI responsible, explainable, and trustworthy to humans. AI ethics entered a phase to make AI “human-centric.”

During this phase, heavily cited papers focused on the application of ethics frameworks and guidelines. In 2020, Rességuier and Rodrigues criticized the prevalent “ethical principles” approach, asserting that ethics should be a continuously refreshed and agile attention to reality. (Rességuier & Rodrigues, 2020). In 2021, Mark Ryan and Bernd Stahl observed “a large degree of convergence in terms of the principles” in AI ethics. The authors compiled AI ethics guidelines directed toward developers and organizational users (Ryan & Stahl, 2021). In 2022, Ashok et al. identified 14 digital ethics implications for AI across various technologies, pushing the discussion toward ethics implementation (Ashok et al., 2022). In 2023, Morley et al. advanced AI ethics discussion by focusing on the human factors, including their understanding, motivation, and barriers (Morley et al., 2023). Illia et al. (2023) addressed issues related to AI text agents post the emergence of ChatGPT.

These three phases constitute the progression of AI ethics based on bibliometric analysis. The envisioned future of AI ethics involves the development of disciplines to make AI human-centric, the codification of AI ethics categories for reference, the creation of an AI ethics value calculator to quantify the benefits of ethics work, and the establishment of flexible tools for the implementability of AI ethics. As AI deployment intersects with human realities, an array of AI ethics issues and principles is expected to emerge.

## KEY AI ETHICS ISSUES

AI ethics shapes the trajectory of crucial AI discussions profoundly. In this section, we highlight seven pivotal debates that remain inconclusive. While our selection is deliberate, we acknowledge the potential omission of other pertinent topics due to space constraints.

### Collingridge Dilemma

The Collingridge Dilemma, as succinctly articulated by Collingridge, encapsulates a predicament that resonates deeply with AI ethicists: “When change is easy, the need for it cannot be foreseen; when the need for change is apparent, change has become expensive, difficult, and time-consuming”. This dilemma underscores the formidable challenge faced by AI ethicists as they navigate the evolving landscape of artificial intelligence.

In its nascent stages, AI development is characterized by uncertainty, and the potential for unforeseen disruptions looms large. The fear is that, by the time the necessity for change becomes evident, the changes required may be irreversible (Mayne & Collingridge, 1982). AI ethics is entrusted with the responsibility of instituting safeguards against such unforeseen consequences, yet it has been characterized as “toothless” by Rességuier and Rodrigues (2020). The disparity between the pace of AI ethical and regulatory development and the rapid advancement of AI technology places ethicists in a precarious position, akin to navigating a highway without clear traffic signs.

Technology companies, propelled by the momentum of innovation and the prospects of profit margin, often engage in accelerated development with limited consideration for ethical and regulatory frameworks. The consequence is a landscape where discussions on ethics and regulations occur post facto, and the cost of implementing changes becomes prohibitively high (Rességuier & Rodrigues, 2020).

**Table 1. Summary of key AI ethics issues**

| Seq. | Issues | Descriptions | Potential Solutions | Papers |
|---|---|---|---|---|
| 1 | Collingridge Dilemma | As Collingridge stated: "When change is easy, the need for it cannot be foreseen; when the need for change is apparent, change has become expensive, difficult, and time-consuming." That is the exact dilemma AI ethics is facing. | ● Precautionary Principle<br>● Intelligent Trial and Error | Mayne & Collingridge, 1982<br>Rességuier, 2020 |
| 2 | AI Status | Is AI an ethical-impact agent, implicit ethical agent, explicit ethical agent, or full ethical agent? | ● From perpetual slaves to full legal personhood | Matthias, 2004<br>Moor, 2006<br>Sparrow, 2007<br>Bryson, 2010<br>Pagallo, 2018<br>Danaher, 2020<br>Mosakas, 2021 |
| 3 | AI Transparency and Explainability | The challenge of balancing full explicability against acceptable yet partial transparency and explainability. | ● Algorithm-based approach<br>● Trust-based approach,<br>● Use decentralized blockchain | Ackerman, 2017<br>Wachter, 2017<br>Robbins, 2019<br>Bertino, 2019<br>Floridi et al., 2018<br>Kim, 2022<br>Langer 2023 |
| 4 | Privacy Protection in the Age of AI | Privacy is being eroded from within and with unprecedented magnitude, scope, and profundity. | ● Ontological approach<br>● Responsible research and innovation<br>● AI-based maturity models | Smolan, 2016<br>Lobo, 2023<br>Floridi, 2006a<br>Stahl, 2018<br>Schuster, 2022 |
| 5 | Justice and Fairness | The issue of injustice, unfairness, and bias cascaded from deep-rooted cultural and historical realities. | ● Assessment tool<br>● Contextualized Embedding Association Test | Sweeney, 2013<br>Buolamwini, 2018<br>Veale, 2017<br>Li, 2021<br>Pandey, 2021<br>Daniels, 2019<br>Guo, 2021 |
| 6 | Algocracy and Human Enfeeblement | The growing influence of AI and algorithm undermines democracy and marginalizes the majority of humans, leading to "human enfeeblement." | ● Universal Basic Income<br>● Active intervention | Aneesh, 2002<br>Vallor, 2015<br>Danaher, 2016<br>Patwardhan, 2023<br>Jensen, 2023<br>Nolan, 2023 |
| 7 | Superintelligence | Unknown consequences and potential existence risks of creating superintelligence. | ● Build friendly AI<br>● Stop development | Good, 1966<br>Kurzweil, 2005<br>Chalmers, 2010<br>Bostroms, 2014<br>Yampolskiy, 2015<br>Floridi, 2019<br>Altman, 2023 |

The Collingridge Dilemma prompts reflection on potential solutions, such as the "Precautionary Principle" and "Intelligent Trial and Error," each presenting its own set of challenges and complexities in implementation.

The Collingridge Dilemma will continue casting a shadow over the AI ethics community as the field grapples with the imperative to guide AI development responsibly in the face of uncertainties that may prove difficult to unwind.

## AI Status

Artificial intelligence (AI), possessing human-like intelligence and humanoid appearance (with Robotics), prompts profound ethical debates regarding its classification—whether it serves as an ethical-impact agent, implicit ethical agent, or explicit ethical agent or whether it attains the status of a full ethical agent (Moor, 2006)? This classification debate has persisted for decades and raises related ethical questions concerning responsibility attribution for AI actions—whether the responsibility rests with the AI itself, its owners, or its programmers. In 2004, Matthias noted a shift in responsibility dynamics due to the emergence of autonomous learning machines based on neural networks, genetic algorithms, and agent architectures. This shift created a "responsibility gap," as the unpredictable future behavior of these machines made it challenging for manufacturers/operators to foresee and be accountable for their actions (Matthias, 2004). Sparrow, in 2007, explored responsibility attribution for war crimes committed by killer robots, highlighting the inadequacy of assigning responsibility to their designers, commanders, or the robots themselves. Consequently, Sparrow deemed it unethical to deploy such systems in warfare (Sparrow, 2007).

In 2010, Bryson argued against humanizing robots, advocating for their perpetual status as slaves to prevent dehumanization and biased resource allocation (Bryson, 2010). In 2018 and 2019, Bryson further discussed constructing AI systems as moral agents or patients, ultimately deeming both possibilities undesirable (Bryson, 2018). Danaher (2020), on the other hand, argued that robots might possess significant moral status, prompting consideration of a duty of "procreative beneficence" towards robots. Mosakas (2021) focused on social robots, asserting that they should not be considered moral entities unless they acquire conscious experience.

The AI status issue remains crucial and potentially disruptive. In October 2017, Sophia, a robot, was granted Saudi Arabian citizenship, becoming the first robot to receive legal personhood in any country. The event underscores the need for policymakers to establish novel forms of accountability and liability in contracts and business law while cautioning against granting AI robots full legal personhood in the foreseeable future (Pagallo, 2018).

## AI Transparency and Explainability

Artificial intelligence (AI) systems operate by optimizing specific tasks through the iterative refinement of an objective function, primarily utilizing supervised and unsupervised learning from large datasets. Unlike human learning, which stems from widely accepted wisdom and ethical standards, AI lacks comparable provisions, resulting in potentially new rules with each neural network simulation. It is hard to explain why the rules can frequently change. Moreover, the quality of AI learning is contingent on the training data, which is subject to low-cost manipulation. Evan Ackerman's (2017) reporting in *IEEE Spectrum* demonstrated that subtle modifications to street signs could deceive machine learning algorithms, demonstrating the fragility of such learning.

As AI assumes a more prominent role in decision-making, there is a growing need to comprehend the underlying logic guiding these decisions. Discussions center on concepts such as "transparency," "accountability," "intelligibility," or being "understandable and interpretable." Floridi et al. (2018) encapsulated these needs under the term "explicability" to understand and hold AI decision-making processes accountable.

However, the technical complexity of explaining the functionality of intricate algorithmic decision-making systems poses a formidable challenge. Some explanations may offer limited meaningful information to data subjects, prompting debates on their practical value. Legal and technical barriers complicate the endeavor to unveil the "black box" of algorithmic decision-making (Wachter et al., 2017). Scott Robbins questioned if there should be an absolute requirement that AI must in all cases be explainable. He posited that insisting on explicability might be unnecessary for low-risk tasks (Robbins, 2019).

The academic discussions have since transitioned to implementation. A "trust-based approach" to explainability has been recommended by Kim and Routledge (2022), while Bertino et al. (2019)

proposed a vision utilizing blockchain technology in a decentralized fashion. Langer and König (2023) contributed strategies to reduce opacity and enhance transparency in algorithm-based human resource management (HRM).

The issue of transparency and explainability stands as a critical concern within AI ethics. Given AI's transformative impact on society, the public's right to understand the decision-making process is paramount. Balancing full explicability against acceptable yet partial transparency and explainability is a delicate endeavor, considering the potential exponential costs associated with providing comprehensive audits. This area warrants further research and academic exploration to offer clarity on the delicate balance between transparency and practicality.

## Privacy Protection in the Age of AI

Privacy protection has conventionally employed an external-focused strategy to thwart external physical or digital intrusions. However, with the widespread integration of personalized AI applications in our ecosystem, privacy is now being eroded from within. The magnitude, scope, and profundity of this loss are unprecedented, resulting in continuous monitoring and tracking of the entire human populace. There is no denying that "In this vast ocean of data, there is a frighteningly complete picture of us" (Smolan, 2016).

Additionally, "the right to be forgotten" is a fundamental principle of the human being. As Lobo et al. (2023) pointed out, "the application of this right is not straightforward: what does erasing mean in the context of a model learned from data?" Does it mean the deletion of the source data? How about the learned model? AI collects, synthesizes, and learns from vast amounts of data from humans. There are currently no tools to monitor, govern, and audit how and where such information is used, let alone the data and model divestiture.

Various AI ethicists have contributed to this discussion. Floridi (2006a) addressed challenges to theories of informational privacy, emphasizing the efficacy of the ontological theory. Stahl et al. (2018) suggested responsible research and innovation (RRI) as a framing concept for socially acceptable technologies. Schuster (2022) proposed the deployment of AI-based maturity models to enhance privacy protection.

## Justice and Fairness

Justice and fairness have become focal points in AI ethics. Some examples include Sweeney's (2013) finding of discrimination in online ad delivery (Sweeney, 2013), Buolamwini and Gebru's (2018) finding of bias in facial recognition systems that prefer lighter skin colors, and Veale and Binns' (2017) finding of racial and social bias in using the location of a person's residence to derive ethnicity or socioeconomic status. Li et al. (2021) found advantaged users (active) who only account for a small proportion of data enjoy much higher recommendation quality than those disadvantaged users (inactive) in the recommender system. Pandey and Caliskan (2021) found a disparate impact of price discrimination algorithms used by ride-hailing applications.

To combat these issues, Daniels et al. (2019) proposed to develop a racial literacy assessment tool to mitigate machine bias. Guo and Caliskan (2021) introduced the Contextualized Embedding Association Test (CEAT) which can summarize the magnitude of overall bias in neural language models by incorporating a random-effects model.

The persistent attention to justice, fairness, and bias indicates that these issues will remain at the forefront. Given the deep-rooted nature of human bias within cultural and historical contexts, substantial efforts are required before AI can achieve freedom from bias and ensure justice and fairness for all.

## Algocracy and Human Enfeeblement

The term "algocracy," coined by Aneesh of Stanford University in 2002, describes a society run by algorithms, with implications for the moral and political legitimacy of decision-making processes (Aneesh, 2002). In 2015, Vallor explored moral deskilling and upskilling in a new machine age, cautioning close attention and perhaps active intervention (Vallor, 2015). Danaher further explores the challenges posed by algorithmic governance, potentially undermining democracy and risking autocracy (Danaher, 2016). The growing influence of AI and robotics introduces ethical challenges, particularly as jobs are automated, and power becomes concentrated, potentially marginalizing the majority of humans and leading to "human enfeeblement," as highlighted by Patwardhan, who noted "perhaps the greatest threat of AI is the potential for loss of meaning in life and human-technology-created enfeeblement in a large section of humanity" (2023).

The excitement of technologists and entrepreneurs contrasts with the confusion among policymakers and the apprehension of the working class. Instances such as protests by San Francisco taxi drivers in 2023 against Cruise driverless cars highlight the concerns surrounding automation (Jensen, 2023). Even creative workers, like Hollywood writers and actors, expressed grievances through strikes in 2023, addressing issues related to AI's role in the creative process (Flint, 2023). A Goldman Sachs report estimates that AI systems, including ChatGPT, could impact 300 million full-time jobs globally, with administrative and legal roles facing the highest risk (Nolan, 2023).

The current debate shifts from concerns about self-realization through fulfilling jobs to grappling with AI's intrusion. Universal Basic Income, a 500-year-old policy idea from English statesman and philosopher, Thomas More, gained renewed interest. As society navigates the intersection of job loss due to automation and the rise of algocracy, there is a pressing need to implement precautions that prevent human enfeeblement in this fast-changing landscape.

## Superintelligence

Superintelligence, also known as the singularity, refers to a hypothetical AI, robot, or agent surpassing the intellectual capabilities of the brightest humans. The ethical discourse revolves around the question of whether humans should proceed with creating an entity beyond human intelligence or refrain from doing so before potential risks emerge.

In a seminal 1966 article titled "Speculations Concerning the First Ultraintelligent Machine," I. J. Good envisioned an ultraintelligent machine capable of surpassing all intellectual activities of humans, emphasizing that it could design even better machines. Good proposed that this ultraintelligent machine would be humanity's last invention (Good, 1966) (Chalmers, 2010). Ray Kurzweil, in his 2005 book *The Singularity Is Near*, predicted the emergence of the singularity, a future where human bodies and brains merge with machines, resulting in nonbiological human intelligence trillions of times more powerful. Philosopher Nick Bostrom, in 2014, cautioned against the development of superintelligent machines, highlighting potential catastrophic risks to humanity if not managed properly. Bostrom emphasized the need to build friendly AI to mitigate these risks (Bostroms, 2014). While Floridi expressed skepticism about doomsday scenarios associated with superintelligence (Floridi, 2019), influential figures such as Elon Musk and late Stephen Hawking took the idea seriously. In 2023, OpenAI leaders published governance recommendations for superintelligence, foreseeing its potential realization within a decade (Altman, 2023).

The prospect of superintelligence raises concerns about the asymmetrical power dynamic between humans and superintelligent entities. Yampolskiy, in "Artificial Superintelligence: A Futuristic Approach," stressed the importance of addressing potential consequences, ranging from economic challenges to humanity's extinction (Yampolskiy, 2015). Even if the risks are deemed low, the consequences of superintelligence, if realized, could be profoundly serious. AI ethicists may need to develop a framework to preemptively address the ethical implications and prevent the unintended development of superintelligence.

Table 2 summarizes the seven key issues. As AI continues its rapid development, new issues will emerge, and old issues may evolve in different directions.

## KEY GAPS

Throughout our review, we identified notable gaps in the field of AI ethics. In this section, we will elaborate on two significant gaps: large ethics model (LEM) and AI identification.

### Large Ethics Model (LEM)

We have introduced the term large ethics model (LEM) as a counterpart to the well-known large language model (LLM), exemplified by systems like ChatGPT and Bard. The methodologies employed in LLMs, renowned for their versatility in handling diverse languages, can be extrapolated to the realm of AI ethics. However, scholarly exploration in this particular field remains minimal.

The profound connection between language and ethics in human communication and behavior is noteworthy. Both serve as frameworks for expressing values, beliefs, and cultural norms, contributing to the establishment of shared understanding and social order. While ethics provides the normative framework guiding decision-making, language acts as a vital vehicle for articulating, disseminating, and debating ethical concepts. Possessing millennia of history, both language and ethics undergo evolution and exhibit cultural and geographical variations.

Presently, AI ethics predominantly follows a conventional approach, commencing with theories, principles, and frameworks—reminiscent of the early days of language translation. In contrast, LLMs utilize deep learning, proving notably efficient in handling intricate subjects like language. Given the success of LLMs in diverse linguistic contexts, there is potential to apply a similar approach to navigate the complexities of different ethical frameworks. We extend an invitation for increased scholarly research in this promising space.

### AI Identification

While the status of AI has garnered considerable attention, the matter of AI identification remains relatively understudied. AI identification involves the labeling of distinct AI instances, akin to industry-standard software license numbers and hardware serial numbers. However, as of now, there is no standardized system for AI identification. To the general public, AI is often perceived as a vague concept rather than a specific product. As AI becomes ubiquitous in human society, the introduction of an AI ID would facilitate the identification, categorization, and expectation-setting for users, offering necessary precautions before interacting with the product.

Facilitating AI identification necessitates the implementation of related supporting services:

- Central Registry: A centralized registry accommodating all AI products, irrespective of their public-facing or non-public-facing nature. This registry draws an analogy to the issuance of birth certificates upon release and death certificates upon decommission. An openly accessible registry, leveraging blockchain technology, can potentially be a viable option.
- Voluntary Rating System: Similar to the Motion Picture Association of America's (MPAA) movie rating system (e.g., G, PG, PG-13), a voluntary rating system for AI products could categorize them based on suitable age range, intelligence level, and creativity for various user demographics.

Once AI entities are assigned IDs, AI ethicists can evaluate and assign ethical scores, ensuring that consumers and users are informed about the nature and origin of the AI with which they interact. The above are preliminary discussions. We would invite more scholarly exploration into AI identification.

## LIMITATIONS AND CONCLUSION

This review is limited by several factors:

- The literature search is based on SCOPUS. Articles not listed in SCOPUS would be excluded. In addition, the literature search was based on title and keywords. Some papers might be missing due to the configuration of the search terms.
- More than 97% of all literature included in the analysis was in English. There is a very high probability that we missed important articles published in other languages.
- The selection of key AI discussions is not entirely scientific: There is a plethora of AI ethics issues being discussed, some directional, some technical, and some operational. In selecting the key discussions, we focused on directional and strategic topics. The authors' discretions may have resulted in important topics being excluded.

In conclusion, our bibliometric analysis based on 20 years of AI ethics literature has revealed three phases in AI ethics development, namely incubation, making AI human-like machines, and making AI human-centric machines. Based on key literature reviews, we introduced seven important yet unconcluded AI ethical issues. In the gap section, we invite more scholarly discussions on the large ethics model and AI identification.

We believe AI ethicists need to get ahead of AI technology to develop ethics disciplines to make AI human-centric, codify an AI ethics category to make AI ethics referenceable, create an AI ethics value calculator to quantify the benefit of ethics work and build flexible tools to make AI ethics implementable.

## ACKNOWLEDGMENT

Special thanks to Erica Baranski from California State University, East Bay for reviewing and editing.

## COMPETING INTERESTS

The authors of this publication declare there are no competing interests.

## FUNDING

The work reported herein was supported by the National Science Foundation (NSF) (Award #2246920). Any opinions, findings, conclusions, or recommendations expressed in this material are those of the authors and do not necessarily reflect the views of the NSF.

*Kevin Gao is an adjunct faculty member in the Management Department at California State University, East Bay (CSUEB). His research interests center around artificial intelligence, robotics, large language models (LLM), and the ethical and regulatory aspects associated with these fields. He has over 20 years of industry experience in consulting, business strategy, and product strategy. He holds a master's degree from the Stanford Graduate School of Business and a bachelor's degree from the School of Management at Shanghai Jiaotong University.*

*Andrew Robert Haverly received his BS and MS in Computer Engineering from Rochester Institute of Technology. He is currently pursuing his Ph.D. in Computer Science & Engineering at Mississippi State University. Andrew is particularly interested in nanorobotics, quantum computing, and artificial intelligence.*

*Sudip Mittal is an Assistant Professor in the Department of Computer Science & Engineering at Mississippi State University. His research interests fall broadly in the areas of Cybersecurity, Cyber-Physical Systems, and Artificial Intelligence. More specifically, his work focuses on building self-protecting systems, autonomous intrusion response, along with predictive maintenance and security of unmanned vehicles/aircraft. Dr. Mittal has published over 70 journals and conference papers in leading venues. Mittal's work has been referenced in The LA times, Business Insider, WIRED, and The CyberWire. He is a member of the ACM and IEEE.*

*Jiming Wu is a Full Professor at California State University, East Bay. He received his B.S. from Shanghai Jiao Tong University, M.S. from Texas Tech University, and Ph.D. from the University of Kentucky. His research interests include Artificial Intelligence, Big Data analytics, and IT adoption and acceptance. His work has appeared in MIS Quarterly, Journal of the Association for Information Systems, European Journal of Information Systems, Information & Management, Decision Support Systems, and elsewhere.*

*Jingdao Chen is an Assistant Professor in the Department of Computer Science and Engineering at Mississippi State University. He received his B.S. degree in Electrical Engineering in 2015 from Washington University in St. Louis and his Masters in Computer Science and Ph.D. in Robotics degrees from Georgia Institute of Technology in 2019 and 2021, respectively. His research lies at the intersection of AI, computer vision, and robotics, and he is interested in exploring data-driven deep learning techniques for scene understanding to enable mobile robots to autonomously navigate and explore unknown environments.*